\documentclass[12pt,a4paper]{article}
\pdfoutput=1
\usepackage[utf8]{inputenc}
\usepackage[T1]{fontenc}
\usepackage{style}
\usepackage[english]{babel}
\usepackage{url}
\usepackage{footnote}

\usepackage{amsmath}
\usepackage{graphicx} 
\usepackage{verbatim} 

\newcommand{\be}{\begin{equation}}
\newcommand{\ee}{\end{equation}}
\newcommand{\beqa}{\begin{eqnarray}}
\newcommand{\eeqa}{\end{eqnarray}}

\newcommand\hMpc{h\text{Mpc}^{-1}}

\renewcommand\k{{\bf k}}

\newcommand{\bseq}{\begin{subequations}}
\newcommand{\eseq}{\end{subequations}}

\renewcommand{\ln}{\mathop{\rm ln}\nolimits}

\renewcommand{\L}{\Lambda}

\renewcommand{\k}{{\bf k}}
\newcommand{\z}{{\bf z}}

\newcommand{\hmpc}{h{\rm Mpc}^{-1}}
\newcommand{\kmax}{k_{\rm max}}
\newcommand{\lmax}{\ell_{\rm max}}
\newcommand{\astfootnote}[1]{%
\let\oldthefootnote=\thefootnote%
\setcounter{footnote}{0}%
\renewcommand{\thefootnote}{\fnsymbol{footnote}}%
\footnote{#1}%
\let\thefootnote=\oldthefootnote%
}

\relax

\title{
Cosmological constraints without nonlinear redshift-space distortions
}

\author[a]{Mikhail M. Ivanov
\footnote{\texttt{ivanov@ias.edu}}
\astfootnote{Einstein Fellow}
}
\author[a,b,c]{Oliver H.\,E. Philcox
}
\author[d]{Marko Simonovi\'c
} 
\author[a]{Matias Zaldarriaga
}

\author[e,f]{Takahiro Nishimichi}
\author[f]{Masahiro Takada}

\affiliation[a]{School of Natural Sciences, Institute for Advanced Study,\\1 Einstein Drive, Princeton, NJ 08540, USA}
\affiliation[b]{Department of Astrophysical Sciences, Princeton University,\\ Princeton, NJ 08540, USA}%
\affiliation[c]{Department of Applied Mathematics and Theoretical Physics, University of Cambridge,\\ Cambridge CB3 0WA, UK}%
\affiliation[d]{Theoretical Physics Department, CERN,\\1 Esplanade des Particules, Geneva 23, CH-1211, Switzerland}
\affiliation[e]{Center for Gravitational Physics, \\ Yukawa Institute for Theoretical Physics, Kyoto University, Kyoto 606-8502, Japan}
\affiliation[f]{Kavli Institute for the Physics and Mathematics of the Universe (WPI), UTIAS \\The University of Tokyo, Kashiwa, Chiba 277-8583, Japan}

\abstract{
Non-linear redshift-space distortions (``fingers of God'') are challenging to model analytically, a fact that limits the applicability of perturbation theory in redshift space as compared to real space. 
We show how this problem can be mitigated using a new 
observable, $Q_0$, which can be easily estimated from the redshift space clustering data and is approximately equal to the real space power spectrum.
The new statistic does not suffer from fingers of God and can be accurately described with perturbation theory down to $k_{\rm max}\simeq 0.4~h~\text{Mpc}^{-1}$. It can be straightforwardly included in the likelihood at negligible additional computational cost, and yields noticeable improvements on cosmological parameters compared to standard power spectrum multipole analyses.
Using both
simulations and observational data from the
Baryon Oscillation Spectroscopic Survey, we
show that improvements vary from $10\%$ to $100\%$ depending on the cosmological parameter considered, the galaxy sample and the survey volume.
}

\begin{document}

\begin{flushright}
YITP-21-106,
	CERN-TH-2021-143
\end{flushright}

\maketitle
\flushbottom

\section{Introduction}

Reliable theoretical models for the intermediate- and short-scale galaxy power spectrum provide the key to obtaining tight constraints on cosmological parameters from current and future spectroscopic galaxy surveys~\cite{Beutler:2016arn,Chudaykin:2019ock,Ivanov:2019pdj,Ivanov:2019hqk,Colas:2019ret,Philcox:2020vvt,Philcox:2020xbv,Sailer:2021yzm}.
In the analysis of the most recent Baryon Oscillation Spectroscopic Survey (BOSS) 
based on the Luminous Red Galaxy (LRG) sample~\cite{Alam:2016hwk},
the main limiting factor in pushing to small scales is the non-linear redshift-space distortions, also known as the ``fingers of God'' (FoG)~\cite{Jackson:2008yv}.
These non-linear effects contaminate the observed galaxy distribution along the line of sight $\hat{\z}$, even on relatively large scales. Further complications come from using the usual multipole expansion of the anisotropic redshift-space power spectrum, since it mixes modes that are parallel and perpendicular to~$\hat{\z}$. As a result, FoG, which affect only the modes along the line of sight, leak into all power spectrum multipoles, significantly limiting the range of scales over which accurate modeling is possible. 

In order to estimate the impact of FoG, it is instructive to compare the range of validity of a given power spectrum model in real and redshift space. Recent analyses of the realistic mock catalogs simulating the BOSS galaxy sample show that the one-loop redshift space
perturbation theory model breaks down at~$k_{\rm max}\simeq 0.25~\hMpc$~\cite{Nishimichi:2020tvu,Ivanov:2019pdj,DAmico:2019fhj}. On the other hand, the real space data for the same volume can be well described by the one-loop model up to significantly smaller scales, $k_{\rm max}\simeq 0.4~\hMpc$~\cite{Schmittfull:2018yuk,Chudaykin:2020aoj}. A similar picture was observed in
the context of Lagrangian perturbation theory in Refs.~\cite{Vlah:2015sea,Vlah:2018ygt,Chen:2020fxs,Chen:2020zjt}.
While these results depend on the survey volume, the effective redshift, and the type of tracers observed, they suggest that there is a potential to improve measurements of cosmological parameters by isolating FoG and extracting the information 
from the transverse Fourier modes (perpendicular to $\hat{\z}$) that are not affected by the non-linear redshift-space distortions.

Throughout the years, many methods had been proposed in order to achieve this goal. The most intuitive approach is to use the redshift-space power spectrum wedges~\cite{Kazin:2011xt,Grieb:2016uuo}. In Fourier space, such techniques effectively operate at the level of the anisotropic power spectrum $P(k,\mu)$, where $\mu\equiv \hat \k \cdot \hat\z$, 
and allow one to use a $\mu-$dependent $k_{\rm max}$ in the analysis~\cite{Grieb:2016uuo}. While conceptually simple, the main shortcoming of wedges is that they cannot be efficiently estimated using FFT techniques, and in practice one has to estimate ``pseudo wedges'', obtained from the standard power spectrum multipoles~\cite{Grieb:2016uuo}. Alternatively, several prescriptions have been used to ``remove'' FoGs directly at the map level~\cite{Tegmark:2001jh,SDSS:2003tbn,Reid:2008zu}, but it remains unclear if the additional systematic errors produced by such methods produce are too large for current and upcoming spectroscopic surveys~\cite{Reid:2008zu}.

In this paper, we build upon ideas from older works~\cite{Hamilton:2000du,Tegmark:2001jh,SDSS:2003tbn,Scoccimarro:2004tg} and use a simple alternative statistic, dubbed $Q_0$. This is closely related to the real space power spectrum, and achieves the goal of isolating the FoG. In essence, this is obtained by measuring a particular linear combination of the first few power spectrum multipoles. The main advantages of $Q_0$ are the following: (a) It can be easily measured using conventional power spectrum multipole estimators; (b) Modulo small effects induced by the broadening of the baryon acoustic oscillation (BAO) peak that affect only the BAO wiggles, $Q_0$ is equal to the real space power spectrum, and can be modeled to higher $k_{\rm max}$; and (c) Its covariance matrix can be straightforwardly computed either analytically or from mock catalogs. $Q_0$ can thus be easily included in the galaxy power spectrum likelihood at negligible extra cost, opening up the possibility to partially include additional small-scale information and improve cosmological constraints compared to conventional power spectrum multipole analyses.

Before we dive into the details, it is worth pointing out the main difference in our approach compared to all previous work, which is related to reliably estimating the covariance for $Q_0$. The problem 
arises from the fact that the non-linear clustering generates all possible multipoles, 
whose covariance rapidly increases with the multipole order, $\ell$. Therefore, if one attempts to produce a better estimate of the real space power spectrum using information from higher and higher $\ell$, the estimator quickly becomes very noisy, and essentially contains no information. This is clearly a paradox. In this work, we show how to resolve this issue and estimate $Q_0$ in a systematic fashion, while keeping the covariance under control. Our method is based on the theoretical error covariance approach~\cite{Baldauf:2016sjb,Chudaykin:2020hbf}. The key idea is to impose natural priors on the smoothness of the higher order multipoles, which, as we will show, effectively suppresses their contribution to the covariance of $Q_0$, while still contributing to the statistic itself. This approach allows multipoles up to arbitrary $\ell_{\rm max}$ to be included in the analysis if needed, guaranteeing the optimal error bars on $Q_0$.

Our paper is organized as follows. We begin with a preliminary discussion in Sec.~\ref{sec:prel}, showing how $Q_0$ can be built from the usual Legendre multipoles with $\lmax=4$ and discuss its relation to the real space power spectrum. Our approach is generalized 
to the case of general $\lmax$ in Sec.~\ref{sec:full}. 
Validation on large-volume N-body simulation data is given in Section~\ref{sec:res}, and applications to the real BOSS data and 
the DESI-like mocks are shown in Section~\ref{sec:boss}.
Finally, we draw conclusions 
in Section~\ref{sec:disc}. Some additional material 
is presented in Appendix~\ref{sec:gen}.

Throughout most of this paper, we will use the PT challenge simulation data~\cite{Nishimichi:2020tvu}, comprising BOSS-like mock catalogs with cumulative volume $\sim 566~(\text{Gpc}/h)^3$. We use a combination of ten independent simulation boxes 
with side length $L=3840~$Mpc/$h$ and $3072^3$ particles each.  
For our purposes, we use only a single redshift bin with $z=0.61$. We will describe the data from the mocks using one-loop perturbation theory templates, as implemented in the \texttt{CLASS-PT} code~\cite{Chudaykin:2020aoj}. The parameter constraints are obtained with
the \texttt{Montepython} MCMC sampler \cite{Brinckmann:2018cvx,Audren:2012wb} and analyzed using the \texttt{getdist} package~\cite{Lewis:2019xzd}.

\section{Preliminary analysis}
\label{sec:prel}

It is instructive to begin with 
a simplified example whereupon $P(k,\mu)$ is fully characterized by its
first four moments, just as in linear theory~\cite{Kaiser:1987qv}. 
In this instance, there is a simple rotation-like transformation 
between the moments of $\mu$
and the Legendre multipoles~$P_{\ell}$,
\be
P(k,\mu)=\sum_{\ell=0,2,4}P_\ell(k) {\cal L}_\ell(\mu)=\sum_{n=0,2,4}Q_n(k) \mu^{n}\,,
\ee
where ${\cal L}_\ell$ is the Legendre polynomial of order $\ell$.
The power spectrum perpendicular to the line-of-sight, i.e. at $\mu =0$, is given by $Q_0$. 
By definition, this coincides with the real-space galaxy power spectrum, 
which can be well described by the one-loop PT model up to $k_{\rm max}\sim 0.4~\hMpc$ \cite{Schmittfull:2018yuk}. 
In contrast, the one-loop PT model for 
$Q_2$ and $Q_4$ breaks down on larger scales, since these moments are dominated by FoG~\cite{Ivanov:2019pdj,Nishimichi:2020tvu,Chudaykin:2020aoj}. 
FoG are a strong UV-effect that lowers the cutoff of the redshift-space effective field theory~\cite{Senatore:2014vja,Lewandowski:2015ziq,Perko:2016puo}. 
Indeed, estimates from the BOSS LRG sample 
give a redshift-space cutoff~\cite{Beutler:2016arn},
\be 
k_{\rm NL, FoG}\approx \sigma_v^{-1}\simeq 0.25~\hMpc\quad\,,
\ee
where $\sigma_v$ is the short-scale velocity dispersion. This can be contrasted with the 
the cutoff of the real space effective field theory $k_{\rm NL, rs}$ (see Refs.~\cite{Baumann:2010tm,Carrasco:2012cv,Chudaykin:2020hbf}), 
which~\cite{Baldauf:2016sjb} estimates to be
\be
k_{\rm NL, rs}\simeq 0.5~\hMpc\,.
\ee
Our goal is to extract
the information contained in $Q_0$ while marginalizing over $Q_2$ and $Q_4$.
An important problem is that the quantities measured by
standard FFT power spectrum estimators 
are the multipoles (see e.g.~\cite{Yamamoto:2005dz,Scoccimarro:2015bla,Hand:2017irw})
and not the moments of $\mu$.
The multipoles pick up contributions from all moments, including those affected by FoG,~i.e.
\be
\label{eq:multq}
\begin{split}
& P_0 = Q_0 +\frac{1}{3}Q_2+\frac{1}{5}Q_4\,,\quad \quad P_2 = \frac{2}{3}Q_2+\frac{4}{7}Q_4\,,\quad \quad P_4 = \frac{8}{35}Q_4\,.
\end{split}
\ee
However, given these simple linear relations, one can easily construct 
an estimator for $Q_0$ from the multipole estimators.
Indeed, a straightforward estimator for $Q_0$ is given by the usual Scoccimarro-Yamamoto formula \cite{Scoccimarro:2015bla},
\be
\label{eq:est}
\begin{split}
\check Q_0(k_i)&=\check P_0-\frac{1}{2}\check P_2+\frac{3}{8}\check P_4 \\
&= \frac{1}{V}\int_{k_i}\frac{d^3k}{4\pi k_i^2 \Delta k}
\delta_0\left((2\cdot 0+1)\delta_0-\frac{(2\cdot 2+1)}{2}\delta_2+\frac{3(2\cdot 4+1)}{8}\delta_4\right)\,,
\end{split}
\ee
where $V$ is the survey volume, and $\int_{k_{i}}$ is the integral over the momentum shell of width $\Delta k$ which is centered at $k_i$.
Moreover, we have assumed the flat-sky approximation and the Kaiser limit~\cite{Kaiser:1987qv} 
for the local redshift-space overdensity $\delta_\ell$, weighted with the appropriate Legendre polynomials,
\be 
\delta_\ell(k,\mu) \equiv (b_1+f\mu^2)\delta_{\rm lin}(k){\cal L}_\ell (\mu)\,.
\ee 
In practice, if the measurements of $P_{0,2,4}$ are available, 
$Q_0$ can be constructed from this datavector by a simple linear summation of these multipoles with appropriate coefficients.
The covariance matrix for $Q_0$ can be obtained directly 
from the 
estimator~\eqref{eq:est},
\be 
\label{eq:covar}
\begin{split}
&\langle \check Q_0(k_i)\check Q_0(k_j)\rangle - \langle \check Q_0(k_i)\rangle \langle \check Q_0(k_j)\rangle \\
&= \frac{(2\pi)^3 \delta_{ij}}{V4\pi k_i^2 \Delta k}\int_0^1 d\mu\, P(k_i)^2 (b_1+f\mu^2)^4 \left(
{\cal L}_0(\mu)-\frac{2\cdot 2+1}{2}{\cal L}_2(\mu)+\frac{3}{8}(2\cdot 4+1){\cal L}_4(\mu)
\right)^2\,\\
&=\frac{(2\pi)^3 \delta_{ij}}{V4\pi k_i^2 \Delta k}\left(
\frac{225 \check{P}_0^2}{64}-\frac{225 \check{P}_0 \check{P}_2}{88}+\frac{3775 \check{P}_0 \check{P}_4}{2288}+\frac{6975 \check{P}_2^2}{9152}-\frac{775 \check{P}_2 \check{P}_4}{1144}+\frac{54975 \check{P}_4^2}{155584}
\right)\,.
\end{split}
\ee
Note that $P_0$ in the above formula contains the stochastic shot-noise term, 
equal to the inverse number density $\bar{n}^{-1}$ in the Poisson limit.
The leading contribution to the covariance is given by the monopole moment (including the shot-noise),
\be 
\frac{2}{N_k}\frac{225 P_0^2}{64}\simeq \frac{2}{N_k} 3.5 P_0^2\,,
\ee
which is $3.5$ times larger than the (auto-)covariance on the monopole, 
and $\sim 4$ times larger than the real 
space covariance (the additional increase is due to the Kaiser effect~\cite{Kaiser:1987qv}).
This apparent inflation of the error bars is driven by higher order multipoles $P_2$ and $P_4$,
which are characterized by a large covariance. 
Thus, the large error on the reconstructed transverse moment $Q_0$ is the inevitable price 
of using the noisy Legendre multipoles in the estimator. 

Alternatively, one can obtain the covariance matrix 
for $Q_0$ directly from the covariance matrix of the multipoles
by an orthogonal transformation dictated by Eq.~\eqref{eq:multq}. 
Denoting this transformation as $P_{\ell} = M_{\ell n}Q_n$ (assuming Einstein summation conventions), we obtain
\be
\begin{split}
& \hat{C}^{(Q)}_{00}= [(\hat{M}^{T})_{0\ell} * \hat{C}^{-1}_{\ell \ell'}* \hat{M}_{\ell' 0}]^{-1}
= \hat C_{00}-\hat C_{02}+\frac{3 \hat C_{04}}{4}+\frac{\hat C_{22}}{4}-\frac{3 \hat C_{24}}{8}+\frac{9 \hat C_{44}}{64}\,,
\end{split}
\ee
which reduces to Eq.~\eqref{eq:covar} in the Gaussian approximation.
For a realistic survey, 
the covariance of $Q_0$ can also be estimated from mock 
catalogs with the usual empirical estimator.

\begin{figure*}[h!]
\centering
 \includegraphics[width=0.49\textwidth]{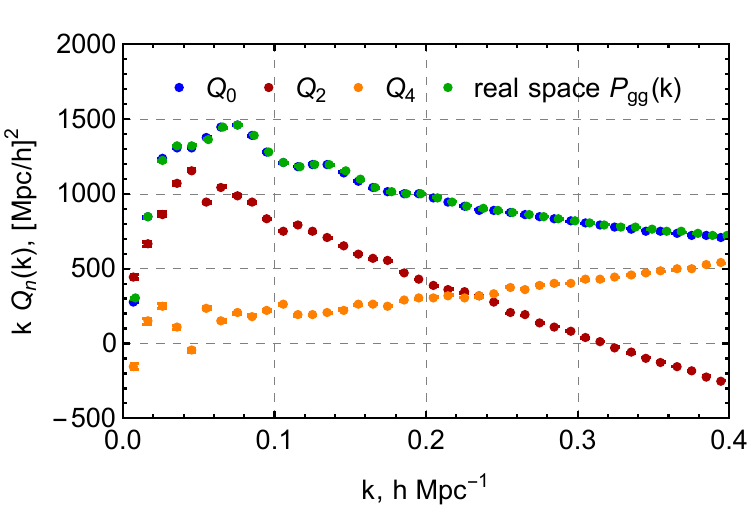}
  \includegraphics[width=0.49\textwidth]{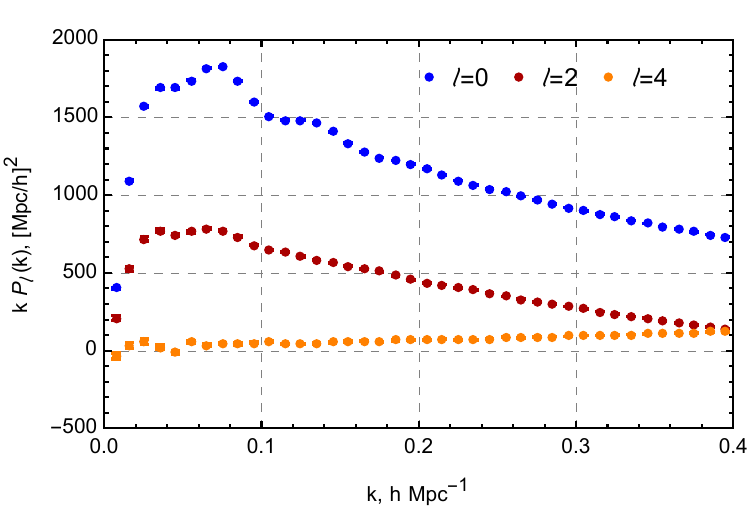}
  \includegraphics[width=0.49\textwidth]{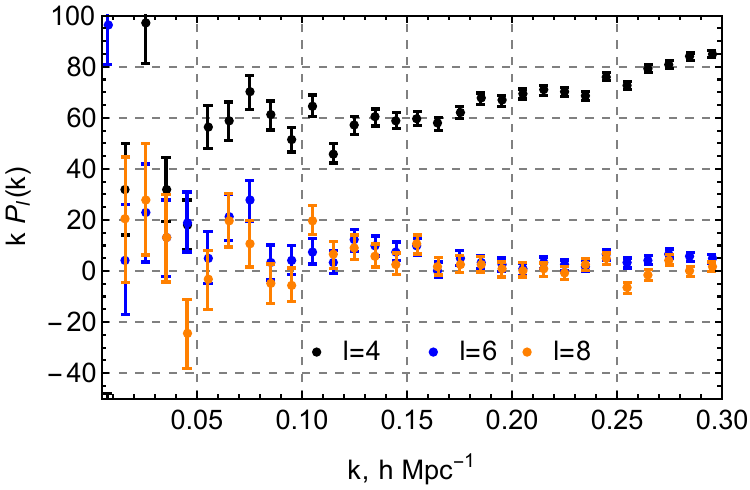}
    \caption{\textbf{Upper panel}: Comparison of moments, $Q_n$, and multipoles, $P_\ell$, for the redshift-space power spectrum of PT challenge galaxies. The real space power spectrum $P_{\rm gg}$ (rescaled to match the AP effect present in $Q_0$) in the left plot is slightly shifted horizontally for clarity, as the datapoints overlap with those of $Q_0$. \textbf{Lower panel}: Higher-order Legendre multipoles with $\ell=4,6,8$.
    }
    \label{fig:qs}
\end{figure*}
Let us consider the $Q_n$ moments extracted from the PT challenge data, as shown in Fig.~\ref{fig:qs}. 
Note that the PT challenge 
redshift space power spectrum moments $P_\ell$ 
are modulated by the Alcock-Paczynski (AP) effect~\cite{Alcock:1979mp}, which is absent in the actual real space power 
spectrum $P_{gg}$, for which the comoving 
distances are computed using the true cosmology.
In order to account for the difference between $Q_0$
and $P_{gg}$, we rescale the latter by the isotropic AP 
factor.
As expected, we see that $Q_0$
is almost identical to the real space power spectrum, once
the AP effect is taken into account (see also Fig.~8 from an earlier work~\cite{Scoccimarro:2004tg}).
However, the higher moments $Q_n$ vary quite significantly on mildly non-linear scales.
In particular, $Q_2$ crosses zero at $k\simeq 0.3~\hmpc$, which may be interpreted
as the PT breakdown for these moments: the zero-crossing means that the non-linear correction
is comparable to the linear one. 
Moreover, non-linearities in the velocity field generate higher-order multipoles with $\ell>4$. 
We show these multipoles (up to $\ell=8$)
estimated from the PT challenge data
in the bottom panel of Fig.~\ref{fig:qs}.
In the presence of higher-order power spectrum multipoles, the estimator for 
$Q_0$ is given by (see Appendix~\ref{sec:gen} for a derivation):
\be 
\check{Q}_0=\check{P_0}-\frac{1}{2}\check{P}_2+\frac{3}{8}\check{P}_4-\frac{5}{16}\check{P}_6+\frac{35}{128}\check{P}_8+...
\ee
In the next section we introduce a general formalism 
that allows one to take higher order multipoles into account consistently.

\section{Formal derivation}
\label{sec:full}

In this section, we will present a general formalism that allows one to 
reconstruct $Q_0$ from any survey for arbitrary $\lmax$. 
We saw in the previous section that using large $\lmax$ in the estimator of $Q_0$ leads to the inflation
of the statistical errors since higher order Legendre multipoles have larger variances. 
However, one can imagine a situation in which the survey volume is such that these moments can become 
important, and their exclusion can lead to noticeable systematic errors. 
To include $Q_0$ for an arbitrary $\ell_{\rm max}$, 
it is more convenient to re-derive the previous results 
using a different approach, which we present here.

\subsection{The case of $\ell_{\rm max}=4$}

Let us start again with the familiar case $\lmax=4$ and consider the likelihood for power spectrum multipoles in the Gaussian diagonal approximation, 
\be
\label{eq:llkl4}
\begin{split}
& -2\ln L(Q_0,Q_2,Q_4) = \Delta \vec{P}_{\ell}
\cdot \hat{C}_{\ell \ell'}^{-1}\cdot \Delta \vec{P}_{\ell'} \,, \quad \text{where}\\
& \Delta \vec{P}_{0}=\left(Q_0(k_i) + \frac{1}{3}Q_2(k_i) + \frac{1}{5}Q_4(k_i)-P_{0}^{{\rm data}}(k_i)\right)\,,\\
& \Delta \vec{P}_{2}=\left(\frac{2}{3}Q_2(k_i) + \frac{4}{7}Q_4(k_i)
-P_{2}^{\rm data}(k_i)
\right)\,,\\
& \Delta \vec{P}_{4}=\left(\frac{8}{35}Q_4(k_i)-P_{4}^{\rm data}(k_i)\right)\,,\\
\end{split} 
\ee
where we have suppressed the explicit summation over multipoles and 
and wavenumber indices.

In the Gaussian approximation all $k$-bins are independent. 
Thus, we can consider the likelihood for each bin separately. 
Marginalizing
the likelihood \eqref{eq:llkl4}
for the $i$-th bin over $Q_2$ and $Q_4$ we obtain we following reduced likelihood:\footnote{For simplicity, we will ignore the logarithmic corrections to the marginalization 
result in 
what follows. 
The leading effect of these corrections is to change the likelihood normalization, which can be neglected in MCMC analysis.
}
\be 
\label{eq:llk_base}
-2\ln L_{\rm marg.}(Q_0) = \sum_{i=1}^{N_{\rm bins}} \frac{(P^{{\rm data}}_{0}(k_i)-\frac{1}{2}P_{2}^{{\rm data}}(k_i)+\frac{3}{8}
P_{4}^{{\rm data}}(k_i)
- Q_0(k_i))^2}{C_{00}-C_{02}+\frac{3 C_{04}}{4}+\frac{C_{22}}{4}-\frac{3 C_{24}}{8}+\frac{9 C_{44}}{64}}\,,
\ee
which exactly coincides with the likelihood for $Q_0$ from the previous section. 
Clearly, this derivation has allowed too much freedom:
we have marginalized over $Q_2$ and $Q_4$ allowing independent and 
arbitrarily large fluctuations in every $k$-bin. 
However, we expect that the scale-dependent FoG contributions are smooth finite functions.
This condition can be implemented by means of the prior on the (unknown) full theoretical model,
along the lines of Ref.~\cite{Baldauf:2016sjb}.

\subsection{Warm-up: theoretical prior on the quadrupole}

Next, let us discuss how the likelihood for $Q_0$ changes if we include some prior information
on the power spectrum multipoles.
Our derivation will closely follow the derivation of the 
covariance matrix in the theoretical error formalism~\cite{Baldauf:2016sjb}.
For simplicity, let us consider a situation in which the redshift-space power spectrum 
depends only on the two moments, $Q_0$ and $Q_2$. 
We need to marginalize this likelihood over $Q_2$. Repeating the derivation above, 
we find following likelihood for $Q_0$ alone:
\be 
-2\ln L(Q_0) = \sum_{i=1}^{N_{\rm bins}}\frac{(P^{\rm data}_{0}(k_i)-P^{\rm data}_{2}(k_i)/2-Q_0(k_i))^2}{C_{00}(k_i) - C_{02}(k_i)+C_{22}(k_i)/4}\,.
\ee

We now assume that there is some 
 prior knowledge of the expectation value $\bar P_{2}$ with some error $E_i$. 
 In other words, there is a likelihood for the 
 theoretical prediction of $\bar P_{2}$, 
 \be
 \begin{split}
 &-2\ln L_{\rm E} = (P_2[Q_2]- \bar P_{2})\cdot \hat{C}_{(P_2)}^{-1} \cdot (P_2[Q_2]- \bar P_{2})= 
(\Delta \vec{Q}_2 - \Delta \vec{Q}'_2)\cdot \hat{\Psi}^{(E)} \cdot  (\Delta \vec{Q}_2 - \Delta \vec{Q}'_2) \,,\\
 & \Delta \vec{Q}_2=\vec{Q}_2- \vec{Q}^{\rm data}_{2}\,,\quad 
 \Delta \vec{Q}'_2=\vec{\bar{Q}}_{2}-\vec{Q}^{\rm data}_2\,,
 \end{split}
 \ee
 where the second equality has rewritten the likelihood for $P_2$
 in terms of the likelihood for $Q_{2}$, using 
 $\bar Q_{2}\equiv 3\bar{P}_2/2$ and defining some precision matrix $\hat{\Psi}^{(E)}$. 
 The split into $\Delta \vec{Q}_2$ and $\Delta \vec{Q}'_2$
 will be clear shortly.
Note that, in principle, the covariance $C_{(P_2)}^{-1}$ is fully correlated. 
The total likelihood takes the following
 form,
\be 
\begin{split}
& -2\ln L(Q_0,Q_2)  = \sum_{i=1}^{N_{\rm bins}} \Delta \vec{P_{\ell}}  C_{\ell \ell'}^{-1}
\Delta \vec{P}_{\ell'}+ \sum_{i,j}^{N_{\rm bins}} C^{-1}_E (P_2[Q_2](k_i) - \bar P_{2}(k_i)) (P_2[Q_2](k_j) - \bar P_{2}(k_j)) \,,\\
&=\Delta \vec{Q}_{m} \cdot \hat{\Psi}_{mn}\cdot \Delta \vec{Q}_{n}
+
\Delta \vec{Q}_2
\cdot \hat{\Psi}^{(E)}\cdot 
\Delta \vec{Q}_2\,.
\end{split}
\ee
The likelihood marginalized over $Q_2$ can be easily obtained, 
\be
\begin{split}
& -2\ln L(Q_0)  = \\
& (\Delta \vec{Q}_0 + 
\{\hat{\Psi}_{00} -\hat{\Psi}_{02}(\hat{\Psi}_{22}+\hat{\Psi}^{(E)})^{-1}\hat{\Psi}_{02}\}^{-1}
(\hat{\Psi}_{02} (\hat{\Psi}_{22}+\hat{\Psi}^{(E)})^{-1} \hat{\Psi}^{(E)}\cdot \Delta \vec{Q}'_2)) \\
& \times 
\left(\hat{\Psi}_{00} -\hat{\Psi}_{02}(\hat{\Psi}_{22}+\hat{\Psi}^{(E)})^{-1}\hat{\Psi}_{02}\right)\\
&\times (\Delta \vec{Q}_0 + 
\{\hat{\Psi}_{00} -\hat{\Psi}_{02}(\hat{\Psi}_{22}+\hat{\Psi}^{(E)})^{-1}\hat{\Psi}_{02}\}^{-1}
(\hat{\Psi}_{02} (\hat{\Psi}_{22}+\hat{\Psi}^{(E)})^{-1} \hat{\Psi}^{(E)}\cdot \Delta \vec{Q}'_2))\,.
\end{split}
\ee
To obtain some insight into the structure of this likelihood, we use several approximations. 
First, let us neglect the cross-covariance $C_{02}$ between the multipoles; this is reasonable since
the normalized correlation coefficient is typically small,
$r_{02}=C_{02}/(C^{1/2}_{00}C^{1/2}_{22})\sim 0.1\ll 1$ for the PT Challenge mocks.
Note that the prediction matrix is not diagonal, i.e.
$\hat \Psi_{02}$ is still non-trivial in this approximation, with
\be
 \hat \Psi_{02} = \hat{C}_{00}^{-1}/3\,.
\ee
As a second approximation, we consider the asymptotic regime $C_E/C \to \infty$. This corresponds to very poor prior knowledge about $\bar P_2$. In this limit, the terms with the theoretical error drop out, and, to leading order in $\mathcal{O}((C_{E}/C)^{-1})$, we obtain
\be 
\label{eq:LLk1}
\begin{split}
&-2\ln L(Q_0)=\\
&=(P^{\rm data}_{0}-P^{\rm data}_{2}/2 - Q_{0})\cdot 
\left(\hat{\Psi}_{00} -\hat{\Psi}_{02} \hat{\Psi}_{22}^{-1}\hat{\Psi}_{02}\right)\cdot (P^{\rm data}_{0}-P^{\rm data}_{2}/2 - Q_{0})\\
&=\sum_{i=1}^{N_{\rm bins}}\frac{(P^{\rm data}_{0}-P^{\rm data}_{2}/2 - Q_{0})^2}{C_{00}+C_{22}/4}\Bigg|_{k_i}
\,,
\end{split}
\ee
where in the last line we have implemented the Gaussian approximation.
Eq.~\eqref{eq:LLk1} gives a usual likelihood with the variance on the estimator $\hat Q_0 = P_0-P_2/2$ reconstructed from the monopole and the quadrupole.
In the opposite limit $C_E/C \to 0$, where the theoretical prior 
is infinitely precise, we have
\be 
-2\ln L (Q_0)=
\sum_{i=1}^{N_{\rm bins}}\frac{( Q_{0} - P^{\rm data}_{0}+\bar P_{2}/2 )^2}{C_{00}}
 + \mathcal{O}(C_E)\,.
\ee 
As expected, at leading order in $C^{(E)}$ adding the prior knowledge on $P_2$ is analogous to fitting the $Q_0$
moment constructed with the prior prediction $\bar P_2$, i.e. 
\be
\hat Q_{0}  = P_0 -\frac{1}{2}\bar P_2\,.
\ee
Importantly, in this case one does not pay the price of including (noisy)
$P_2$ in the estimator for $\hat Q_{0} $, i.e. 
the covariance is given only by the monopole contribution.

\subsection{Generalization to higher order multipoles}

Generalization is straightforward, and takes the form
\be 
\label{eq:masterLLK}
\begin{split}
& -2\ln L(Q_0,...,Q_{\ell_{\rm max}}) = \sum_{i=1}^{N_{\rm bins}} \sum_{\ell,\ell' \leq \ell_{\rm max}}C_{\ell \ell'~{\rm data}}^{-1}\Delta P_{\ell}\Delta P_{\ell'} \\
& + \sum_{\ell=0}^{\ell_{\rm max}}\sum_{i,j}^{N_{\rm bins}} 
(P_\ell[Q](k_i) -  \bar P_{\ell}(k_i)) (\hat{C}^{ ({\rm E}),\,(\ell \ell')})^{-1} _{ij} (P_\ell[Q](k_j) - \bar P_{\ell}(k_j)) \,,
\end{split}
\ee
where $P_\ell[Q]$ denotes a general expression for the 
power spectrum multipole $\ell$ through moments of $\mu$ (see Appendix \ref{sec:gen}).
The prior $\bar P_{\ell}$ can be either a fit to the data with some smooth function, or taken from the perturbation theory prediction. 
For $\ell > 4$ either option  gives an envelope very close to $0$ on mildly non-linear scales.
The final likelihood for $Q_0$ is obtained by marginalizing \eqref{eq:masterLLK}
over all $Q_{\ell}$ functions with $\ell \geq 2$, which can be performed analytically.
In general, smoothness in $\mu$ and $k$ implies that the prior covariance should be $100\%$ correlated~\cite{Chudaykin:2020aoj}:
\be
C^{({\rm E}),~(\ell \ell')}_{ij}= E_\ell(k_i) E_{\ell'}(k_j) \,.
\ee
However, given that the true shape is not known, 
we impose a weaker condition on smoothness in $\mu$ and $k$-space. Namely, we will use~\cite{Chudaykin:2020hbf}
\be
C^{({\rm E}),~(\ell \ell')}_{ij}= E_\ell(k_i) E_{\ell'}(k_j)\, \mathrm{exp}\left(-\frac{(k_i-k_j)^2}{2\Delta k^2}\right)\,\mathrm{exp}\left(-\frac{(\ell - \ell')^2}{2\Delta \ell^2}\right)\,.
\ee
Conservatively, we can choose the theory prior to be 100$\%$ of the expected value, e.g. 
\be
E_\ell(k_i) =  \bar P_\ell (k_i)\,.
\ee
The full likelihood \eqref{eq:masterLLK} simplifies in the two extreme limits, $C_{E} \ll C_{\rm data}$ and $C_{E} \gg C_{\rm data}$. As we have seen in the previous section, in the first case 
one needs to simply replace the true datavector $P_{\ell}$ by $\bar P_{\ell}$ in the estimator of $\hat Q_{0}$. This does not require any change to the covariance; hence, we do not pay the price of using the noisy $P_{\ell}$ in our $Q_0$ estimator.  
In the second case we must use the $P_{\ell}$ from the data when constructing $\hat Q_{0}$
and include the noise of this multipole in our covariance. 
For all practical purposes it is sufficient to work within these two limits. 
To estimate the transition between the two regimes, we may compare the effective 
$\chi^2$ contribution coming from the data and the prior on $P_\ell$:
\be
\begin{split}
& \chi^2_{\rm data}=(\vec{P}_{\ell}-\vec{\bar{P}}_{\ell})
\cdot 
\hat{C}^{-1}_{\ell\ell}
\cdot 
(\vec{P}_{\ell}-\vec{\bar{P}}_{\ell})\\
& \chi^2_{\rm prior}= (\vec{P}_{\ell}-\vec{\bar{P}}_{\ell})
\cdot (\hat{C}^{(E)}_{(\ell \ell')})^{-1}\cdot 
(\vec{P}_{\ell}-\vec{\bar{P}}_{\ell})
\end{split} 
\ee
This suggests the following algorithm to deal with multipoles $\ell \geq 4$:
\begin{enumerate}
	\item Select some $\lmax\geq 4$ and estimate all mutipoles with $\ell \leq \lmax$. Fit these multipoles with some smooth curves $\bar P_\ell$.
	\item For each $4\leq \ell\leq \lmax$ compute the ratio 
	$\chi^2_{\rm data}/\chi^2_{\rm prior}$. 
	If $\chi^2_{\rm data}/\chi^2_{\rm prior}>1$, $P_\ell$ should be included in the $Q_0$ estimator along with its effect on the covariance. In the opposite regime, $\check{Q}_0$ and its covariance should be estimated using only the lower multipoles. For all higher multipoles, add the contribution from the relevant multipole moment $\ell$
	as a smooth prior $\bar P_\ell$  
	to $\check{Q}_0$.
\end{enumerate}

\subsection{Modeling $Q_0$}
Finally, let us discuss the theoretical model for
$Q_0$. In linear theory, $Q_0$ would be the real space galaxy auto power spectrum $P_{\rm gg}$. The situation becomes more complicated 
when IR resummation (i.e.~the effects of long-wavelength displacements, which cannot be treated perturbatively) is taken into account. 
Indeed, the non-linear BAO damping factor is direction-dependent~\cite{Ivanov:2018gjr}. 
Assuming a wiggly/smooth decomposition of the linear power spectrum $P_{\rm lin}=P_{\rm nw}+P_{\rm w}$~\cite{Baldauf:2015xfa,Blas:2015tla,Blas:2016sfa,Vasudevan:2019ewf}, at leading order we may write 
\be
P(k,\mu)=(b_1+f\mu^2)^2\left( P_{\rm nw}+ 
P_{\rm w}e^{-\Sigma^2k^2 (1+f\mu^2(2+f))-\delta \Sigma^2 k^2 f^2 \mu^2(\mu^2-1)}\right)\,,
\ee
where the BAO damping functions are given by 
\be 
\begin{split}
& \Sigma^2 =\frac{1}{6\pi^2}\int_0^{k_S} dq~P_{\rm nw}(q)
(1-j_0(qr_{\rm BAO})+2j_2(qr_{\rm BAO}))\,,\\
& \delta\Sigma^2 = \frac{1}{2\pi^2}\int_0^{k_S} dq~P_{\rm nw}(q)
j_2(qr_{\rm BAO})\,,
\end{split}
\ee
for spherical Bessel functions $j_\ell(x)$, comoving 
BAO scale at the drag epoch $r_{\rm BAO}$, and separation scale $k_S$. 
Since the damping factor is large, the exponential suppression cannot be Taylor-expanded. 
This means, that, in general, 
we need to use an infinite series in $P_\ell$ in our 
estimator of $Q_0$ in order to remove the direction-dependence 
of the BAO wiggles.  
However, as the shape of the BAO wiggles is known analytically to all orders in $\mu$,
we can just compute
redshift-space multipoles, $\{P_\ell\}$, theoretically
and then combine them into $Q_0$ 
just as for the data. 
This guarantees that the suppression of the BAO wiggles 
in $Q_0$ is the same in the datavector and in the theory model.
Thus, our theory model is 
\be
\label{eq:Q0mod}
Q_0(k)=P_0(k)-\frac{1}{2}P_2(k)+\frac{3}{8}P_{4}(k)\,,\\ 
\ee
where $P_{0,2,4}$ contain all necessary redshift-space counterterms. The priors on these counterterms can be extracted from fitting the full datavector $P_{0,2,4}$ at low $k_{\rm max}$, where the perturbative modeling of FoG is still accurate.

\begin{figure*}[h!]
\centering
 \includegraphics[width=0.49\textwidth]{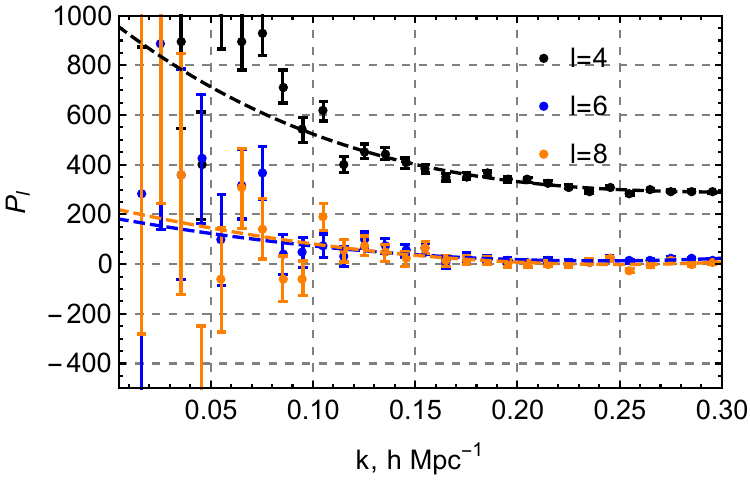}
    \caption{Higher order multipoles of the PT challenge data and their fits by quadratic polynomials.  
    }
    \label{fig:snr}
\end{figure*}

\section{Validation on PT challenge mocks}
\label{sec:res}

In this section we apply the formalism described above to the PT challenge data. 
We will use the Gaussian approximation for all sample covariance matrices, which has been shown to be very accurate for the purpose of parameter constraints~\cite{Wadekar:2020hax}.

\subsection{Estimation of $Q_0$ from the data}

As a first step, we obtain an estimate for $\bar P_\ell$ from the fits to the data, as in the left panel of 
Fig.~\ref{fig:snr}. As a second step, we compute $\chi^2_{\rm data}/\chi^2_{\rm prior}$ assuming the following $100\%$ prior on $P_\ell$:
\be
\begin{split}
C^{\rm (E)~(\ell \ell')}_{ij} = \bar P_\ell(k_i) \bar P_{\ell'}(k_j) \mathrm{exp}\left(-\frac{(k_i-k_j)^2}{2\Delta k^2}\right) \mathrm{exp}\left(-\frac{(\ell - \ell')^2}{2\Delta \ell^2}\right)\,,
\end{split}
\ee
The coherence length $\Delta k$ characterizes
the amount of correlation across different $k$-bins, and ensures that the likelihood properties do not depend on the binning. 
Choosing a large $\Delta k$ increases the significance of the theoretical prior by assuming an extra correlation between
$k$ bins.
The coherence in $\ell$ space 
corresponds to a smoothness of the power spectrum 
as a function of $\mu$.
In practice, we choose
$\Delta k=0.001~h$/Mpc
and assume also that the theoretical prior covariance matrix is diagonal in the multipole space, i.e.~$\Delta \ell = 0$.
This choice corresponds to a very conservative situation where 
the prior on $\bar P_\ell$ is quite poor. Essentially,
we do not require the
two-dimensional power spectrum prior $P(k,\mu)$ 
to be a smooth function in both $k$ and $\mu$.
Even this very conservative situation will be sufficient
for our purposes.
With our choice of $\Delta k$ and $\Delta \ell$, and using $k_{\rm max}=0.3~h/$Mpc, we find
\be
\begin{split}
\frac{\chi^2_{\rm prior}}{\chi^2_{\rm data}}=0.9,~80,~75 \quad \text{for}\quad \ell=4,~6,~8\,.
\end{split} 
\ee
One can also check that ${\chi^2_{\rm prior}}/{\chi^2_{\rm data}}\ll 1$ is always true for $\ell=0,2$.
For a more aggressive choice of $\Delta k=0.01~\hMpc$ we obtain the following numbers:
\be 
\begin{split}
\frac{\chi^2_{\rm prior}}{\chi^2_{\rm data}}=2,~232,~196 \quad \text{for}\quad \ell=4,~6,~8\,,
\end{split} 
\ee
which do not change results qualitatively.
These results suggests that our prior is marginally important for $\ell=4$, and it is much more significant than the actual likelihood contribution for $\ell>4$. If we include off-diagonal-in-$\ell$ matrix elements, which 
correspond to a prior on the smoothness of the power 
spectrum in $\mu$, the significance of 
the priors will increase even further. In particular, for $\Delta \ell=2$, we have:
\be 
\begin{split}
\frac{\chi^2_{\rm prior}}{\chi^2_{\rm data}}=3.3,~152 \quad \text{for}\quad \ell=4,~6\,.
\end{split} 
\ee
On the one hand, the $\ell=4$ prior never
dominates over the data by more than a factor of few, regardless of the set-up.
To be maximally conservative, we will 
always include the hexadecapole in our analysis
and take into account its contribution into the covariance
matrix. On the other hand, we see that the
contribution of higher
order multipoles ($\ell>4)$ is always dominated by priors.
Thus, we conclude that even for the large-volume PT challenge mocks 
the higher-order multipole moments $\ell>4$ can be ignored in the estimation of the covariance matrix for $Q_0$ in the mildly
non-linear regime.
However, we may want to include 
higher multipoles in the form of the mean prior to the theoretical model. 
To this end, we need to check if their inclusion is 
strictly needed to describe the data. 
To that end, we perform several MCMC analyses 
of the $Q_0$ likelihood from the PT challenge data
for different choices of $\lmax$. 

We fit the joint likelihood comprising 
the multipoles $P_{0,2,4}$ for $k_{\rm max}=0.14~\hMpc$ 
and $Q_0$ in the range $0.14~\hMpc\leq k<0.3~\hMpc$. 
Since the $k$-bins do not overlap between the two likelihoods, 
they are uncorrelated in the Gaussian limit. We fit the $P_{0,2,4}$ datavector with the
one-loop effective field theory template of Refs.~\cite{Perko:2016puo,Nishimichi:2020tvu,Ivanov:2019pdj,Chudaykin:2020aoj}. Note that we include the next-to-leading-order operator $\tilde c k^4 \mu^4 P_{\rm lin}(k)$ in our analysis to account for higher-order FoG effects. Additionally, we use
the full set of
stochastic contributions from Refs.~\cite{Perko:2016puo,Schmittfull:2020trd},
\be
\label{eq:stochdef}
P_{\rm stoch}(k,\mu)= \left\{a_0 \left(\frac{k}{k_{\rm NL}}\right)^2 + a_2 \mu^2 \left(\frac{k}{k_{\rm NL}}\right)^2 + P_{\rm shot}\right\}\cdot \frac{1}{\bar n}~[\text{Mpc}/h]^3\,,
\ee 
where $\bar n$ is the inverse number density of tracers.
Using the hexadecapole moment we are able to break the 
strong degeneracy between $a_2$ and $\tilde c$, which is present 
in the $P_{0,2}$ likelihood. As to $Q_0$, we use the model~\eqref{eq:Q0mod}, which depends on the same nuisance parameters 
as our likelihood for the multipoles for $k_{\rm max}=0.14~h$/Mpc. We use the following parameter vector (see~\cite{Chudaykin:2020aoj} for our notations):
\be
\{\omega_m,h,A_s\}\times 
\{b_1,b_2,b_{\mathcal{G}_2},b_{\Gamma_3},c_0,c_2,c_4,\tilde{c},a_0,a_2,P_{\rm shot} \} \,.
\ee
We fix $n_s$ and $\Omega_b/\Omega_m$ to the known fiducial values
as in Ref.~\cite{Nishimichi:2020tvu}. The following Gaussian priors on the nuisance parameters are assumed:
\be
\begin{split}
& a_0 \sim \mathcal{N}(0,1^2)\quad a_2\sim \mathcal{N}(0,1^2)\quad P_{\rm shot}\sim \mathcal{N}(0,0.3^2)\,,\quad b_{\Gamma_3}\sim \mathcal{N}
\left(\frac{23}{42}(b_1-1), 1^2\right)\,, 
\end{split} 
\ee
using flat infinite priors on $b_1, b_2, c_0, c_2, c_4,\tilde{c}$ and $b_{\mathcal{G}_2}$.
The mean value of $b_{\Gamma_3}$ is taken from the prediction of the coevolution model~\cite{Desjacques:2016bnm,Abidi:2018eyd}.
Since the Poissonian shot noise contribution was subtracted from the data, we assume that the 
mean residual contribution is zero, with variance corresponding to $\sim 30\%$ of $\bar{n}^{-1}$,
consistent with the deviations due to the halo exclusions~\cite{Baldauf:2013hka,Schmittfull:2018yuk} 
expected for the BOSS-like host halos.

\begin{figure*}[htb!]
\centering
 \includegraphics[width=\textwidth]{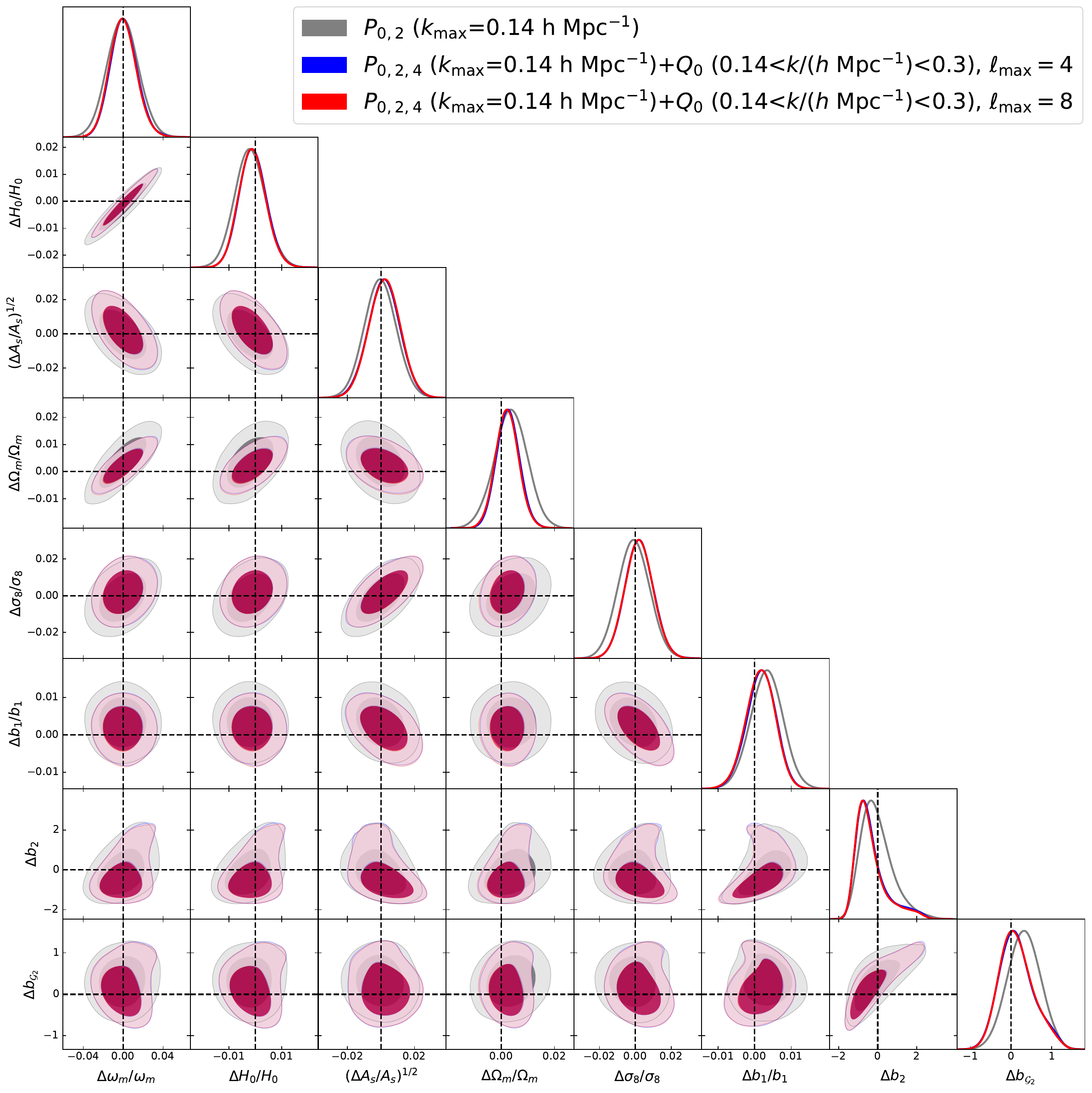}
    \caption{Posteriors from the PT challenge data for the analysis with fixed $\Omega_b/\Omega_m$ and $n_s$.
    }
    \label{fig:mu0}
\end{figure*}

Let us now study the convergence of our method with respect to the value of
the maximal multipolar index. In particular, we consider $\ell_{\rm max}=4,6,8$. 
In all cases, the hexadecapole is fully included in the theory, data and the covariance, with $\ell = 6, 8$ included only via priors.
The results of our analysis are shown in Fig.~\ref{fig:mu0} and Table~\ref{tab:fb}. 
Each parameter $p$ (except $b_2$ and $b_{\mathcal{G}_2}$)
is shown in the format $\Delta p/p\equiv p/p_{\rm true}-1$,
where we use fiducial values for true cosmological parameters, and extract the true value of $b_1$ from the galaxy-matter cross 
spectrum~\cite{Nishimichi:2020tvu}.
For $b_2$ and $b_{\mathcal{G}_2}$ we use the format
$\Delta p \equiv p-p_{\rm true}$, where the true values of $b_2$ and $b_{\mathcal{G}_2}$
are measured from the combined power spectrum and bispectrum
analysis~\cite{Ivanov:2021kcd}.
For comparison, we show also the baseline results for the $P_{0,2}$ likelihood at $k_{\rm max}=0.14~h$/Mpc.
On the one hand, we can see that $Q_0$ narrows the
contours for the $\omega_{m}$ and $H_0$ by $\lesssim 20\%$. 
This improvement is expected since these parameters are measured 
from the shape of $Q_0$.
In contrast, the amplitude parameters $A^{1/2}$
and $\sigma_8$ cannot be accurately measured from the 
real-space galaxy power spectrum alone because of the degeneracy with galaxy bias.
This explains why the posteriors for these parameters do not appreciably shrink 
after the inclusion of $Q_0$.
The priors on higher 
order multipoles in the estimator $\check{Q}_0$ have a negligible effect on the posterior 
distribution, which motivates us to use $\ell_{\rm max}=4$
as our baseline choice.

\begin{table*}[ht!]
\begin{center}
  \begin{tabular}{|c||c|c|c|} \hline
   Parameter   &  $P_{0,2}$  ($\kmax=0.14~\hMpc$)
   &  $P_{\ell}+Q_0,~\lmax=4$  
   &  $P_{\ell}+Q_0,~\lmax=8$
      \\ [0.2cm]
      \hline 
$\Delta H_0/H_0$    & $-0.0020\pm 0.0059  $
& $-0.00096\pm 0.0052$  
& $-0.0011\pm 0.0052 $\\ \hline
$\Delta A^{1/2}/A^{1/2}$   & $-0.0004\pm 0.0096 $
& $0.0019\pm 0.0093 $
& $0.0021\pm 0.0093 $
\\ 
\hline
     $\Delta\omega_{m}/\omega_{m}$  &
$0.000\pm 0.016$
  & $0.001^{+0.013}_{-0.014}$
  & $0.000^{+0.012}_{-0.014}$
   \\  \hline
$\Delta b_1/b_1$   & $0.0033\pm 0.0044   $
& $0.0018^{+0.0041}_{-0.0037} $
& $0.0016\pm 0.0040   $
\\ 
\hline
$\Delta b_2$   & $-0.04^{+0.55}_{-0.96} $
& $-0.33^{+0.31}_{-0.93} $
& $-0.36^{+0.32}_{-0.90}$
\\ 
\hline
$\Delta b_{\mathcal{G}_2}$   & $0.31\pm 0.41  $
& $0.14^{+0.34}_{-0.50}  $
& $0.14^{+0.34}_{-0.49} $
\\ 
   \hline \hline 
$\Delta\Omega_m/\Omega_m$   & $0.0035\pm 0.0062 $
& $0.0025\pm 0.0043 $
& $0.0022\pm 0.0044  $\\ \hline
$\Delta\sigma_8/\sigma_8$   
& $-0.0008\pm 0.0088$
&$0.0021\pm 0.0079$ 
&$0.0021\pm 0.0079 $ \\ 
\hline
\end{tabular}
\caption{Constraint on key cosmological and nuisance parameters from the PT challenge mock power spectra, obtained with fixed $\Omega_b/\Omega_{m}$ and $n_s$ as in Ref.~\cite{Nishimichi:2020tvu}.
$P_\ell$ denotes the datavector $\{P_0,P_2,P_4\}$ with $\kmax=0.14~\hMpc$. The second and third columns 
show results of the addition of $Q_0$ in the range $0.14\leq k/(\hMpc)<0.3$. In the third column we add mean priors on 
the multipole moments with $\ell=6,8$ to the theory model.
Parameters in the upper group part of the table
were varied directly, while the lower group
are the derived parameters.
}
\label{tab:fb}
\end{center}
\end{table*}

\subsection{Cosmological constraints with the $\omega_b$ prior}

\begin{figure}[ht!]
\centering
 \includegraphics[width=\textwidth]{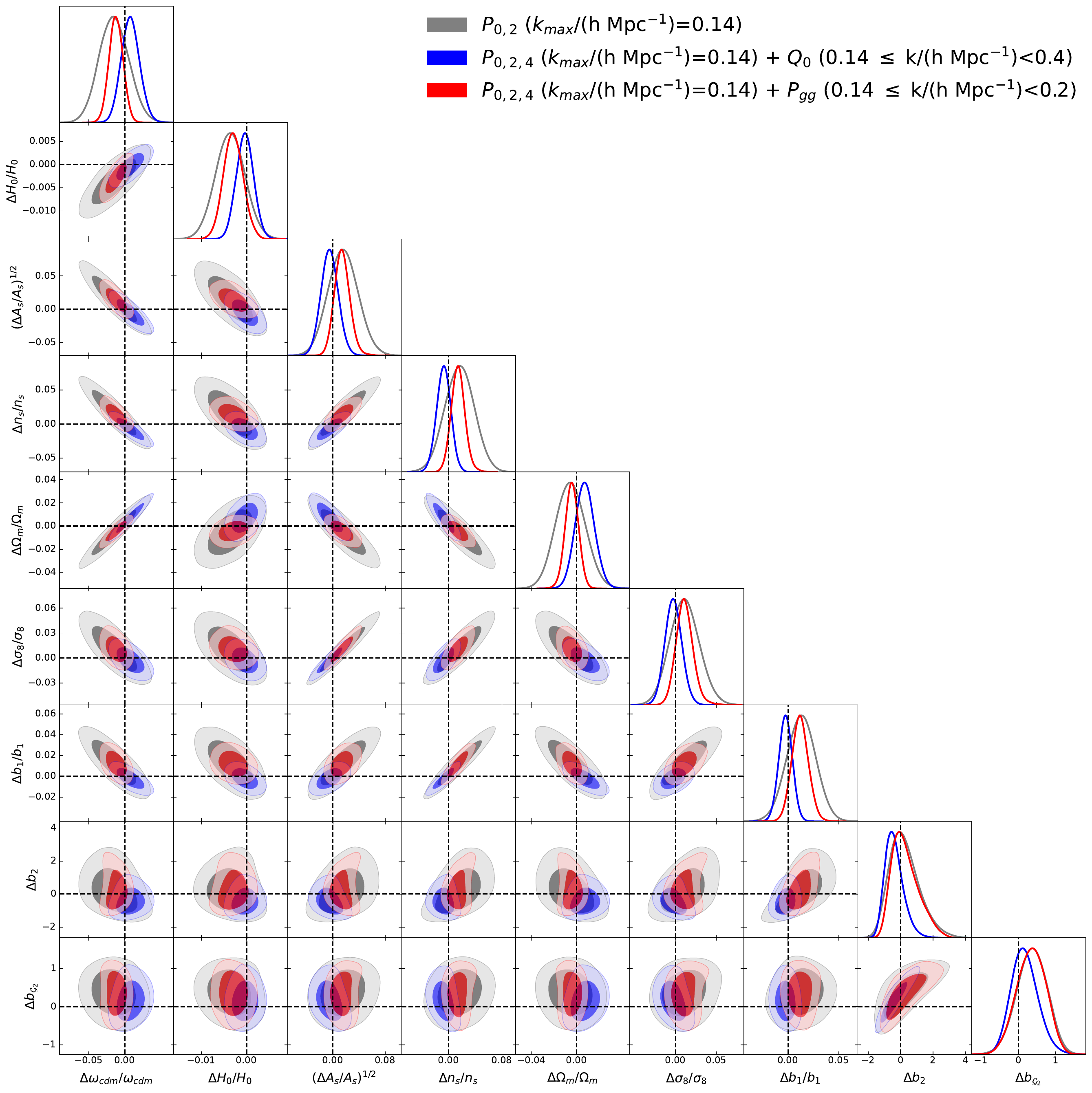}
    \caption{
Posteriors from the analysis of the 
    PT challenge mock galaxy power spectrum
    with a prior on $\omega_b$. 
    }
    \label{fig:mu0ns}
\end{figure}

The information gain from $Q_0$ 
depends on the adopted priors and particular cosmological model. To illustrate this, 
we refitted the mock power spectra 
fixing $\omega_b$ instead of $\Omega_b/\Omega_{m}$,
which was the choice adopted in our previous analysis.
This simulates the addition of the $\omega_b$ prior, which is readily available in e.g. Planck or BBN.
 We additionally allow $n_s$ to vary freely. 
The corresponding results are presented in Fig.~\ref{fig:mu0ns} and in Tab.~\ref{tab:ob}.
 We find that the fit to $Q_0$ is unbiased all the way up to $k_{\rm max}=0.4~\hMpc$.
 We also see that in the case of a single prior on $\omega_b$ the addition of $Q_0$
 improves the constraints on all the remaining cosmological parameters roughly 
 by a factor of 2. 

It is useful to compare our results with the case of the true real space galaxy power spectrum, using the same $k$ ranges.
To that end, we replace $Q_0$ with the actual 
real-space power spectrum $P_{\rm gg}$ extracted from the same PT Challenge simulations, and refit the data.
The diagonal elements of the Gaussian covariance for $P_{\rm gg}$ are roughly four times smaller than similar elements of $Q_0$ for the same volume and shot noise.
As a result of this small covariance, the one-loop perturbation theory fit to $P_{\rm gg}$ becomes 
biased beyond 
$k_{\rm max}=0.2~\hMpc$, which we adopt 
as a baseline data cut in this case.
The resulted parameter limits are very similar to those obtained 
from our baseline $Q_0$ analysis at $k_{\rm max}=0.4~\hMpc$. 
This matches the expectation 
that the two statistics should be equivalent 
at the level of total information 
for appropriate data cuts.

\begin{table*}[ht!]
\begin{center}
  \begin{tabular}{|c||c|c|c|} \hline
   Parameter   &  $P_{0,2},~(k_{\rm max}=0.14)$ &  $P_\ell+Q_0~(k_{\rm max}=0.4)$  
     &$P_\ell+P_{\rm gg}~(k_{\rm max}=0.2)$ 
     \\ [0.2cm]
      \hline 
$\Delta H_0/H_0$    & $-0.0036\pm 0.0032  $
& $-0.0004\pm 0.0019$  
& $-0.0029\pm 0.0022  $\\ \hline
$\Delta A^{1/2}/A^{1/2}$   & $0.016\pm 0.023 $
& $-0.005\pm 0.013  $
& $0.014^{+0.011}_{-0.012} $
\\ 
\hline
     $\Delta\omega_{cdm}/\omega_{cdm}$  &
$-0.015\pm 0.020$
  & $0.008\pm 0.013  $
  & $-0.0121\pm 0.0096$
   \\  \hline
     $\Delta n_s/n_s$  &
$0.016\pm 0.022 $
  & $-0.008\pm 0.011$
  & $0.014\pm 0.010$
   \\  \hline
   $\Delta b_1/b_1$   & $0.013\pm 0.014   $
& $-0.0024\pm 0.0066 $
& $0.0119\pm 0.0082   $
\\ 
\hline
$\Delta b_2$   & $0.25^{+0.70}_{-1.1}$
& $-0.40^{+0.42}_{-0.66} $
& $0.29^{+0.56}_{-1.0} $
\\ 
\hline
$\Delta b_{\mathcal{G}_2}$   & $0.36\pm 0.40  $
& $0.17^{+0.31}_{-0.39}  $
& $0.34^{+0.41}_{-0.37}  $
\\ 
   \hline \hline 
$\Delta\Omega_m/\Omega_m$   & $-0.005\pm 0.013 $
& $0.0073\pm 0.0085 $
& $-0.0044\pm 0.0059 $\\ \hline
$\Delta\sigma_8/\sigma_8$   
& $0.011\pm 0.018$
&$-0.003\pm 0.010 $ 
&$0.0110^{+0.0094}_{-0.011} $ \\ 
\hline
\end{tabular}
\caption{Constraints from the analysis of the PT challenge data with the $\omega_b$ prior. $P_\ell$ denotes the datavector $\{P_0,P_2,P_4\}$ with $\kmax=0.14~\hMpc$. The second column 
shows results of the addition of $Q_0$ in the range $0.14\leq k/(\hMpc)<0.4$; while in the third column instead we add
the actual real space power spectrum $P_{gg}(k)$ in the range 
$0.14\leq k/(\hMpc)<0.2$.
}
\label{tab:ob}
\end{center}
\end{table*}

\section{Applications to realistic surveys}
\label{sec:boss}

So far we have studied $Q_0$ in application
to the PT Challenge mocks whose total volume is 
$566~h^{-3}$Gpc$^3$ at the effective redshift $z=0.61$.
Current and future surveys will have somewhat 
smaller volumes, therefore it is useful to test
to what extent the real space power spectrum can 
improve cosmological parameter measurements from 
realistic surveys. We address this question in this section
and analyze the spectroscopic data from BOSS
and DESI-like mock catalogs. 

\begin{figure}[ht!]
\centering
 \includegraphics[width=\textwidth]{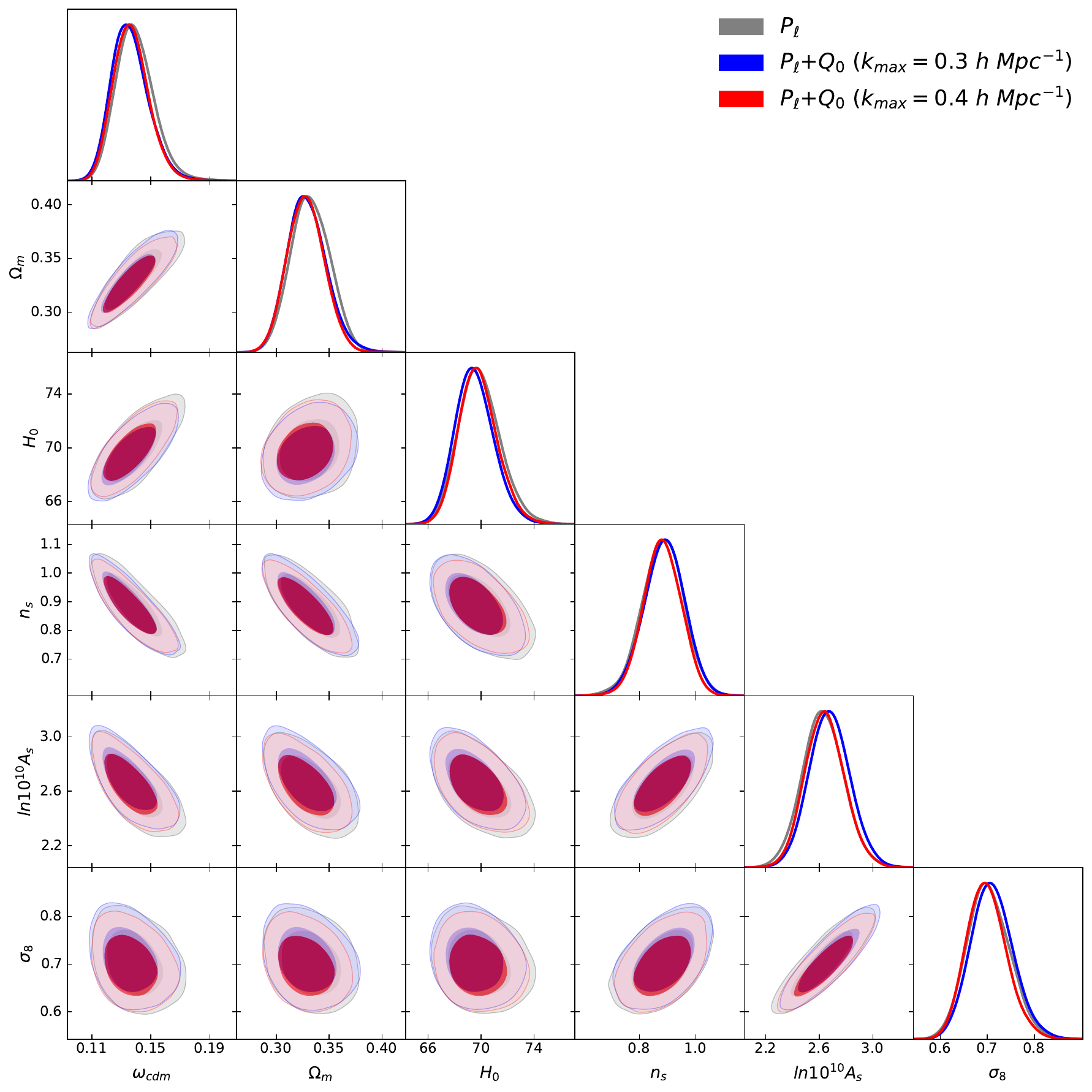}
    \caption{
Posteriors from the cosmological analysis of the 
    BOSS galaxy power spectrum measurements
    combined
    with the BBN prior on $\omega_b$. 
    }
    \label{fig:boss}
\end{figure}
\begin{table*}[ht!]
\begin{center}
  \begin{tabular}{|c||c|c|c|c|} \hline
   Parameter   &  $P_{\ell}$
   &  
   $P_\ell+Q_0$
    $~\left(\frac{k_{\rm max}}{\hMpc}=0.3\right)$  
     &  
   $P_\ell+Q_0$
    $~\left(\frac{k_{\rm max}}{\hMpc}=0.4\right)$  
     \\ [0.2cm]
      \hline 
$H_0$/(km/s/Mpc)    & $69.89_{-1.7}^{+1.5}$
& $69.51_{-1.6}^{+1.3}$ 
& 
$69.79_{-1.6}^{+1.3}$
\\ \hline
$\ln(10^{10}A_s)$   & $2.63_{-0.16}^{+0.15}$
& $2.68_{-0.16}^{+0.15}  $ & $2.64_{-0.16}^{+0.14}$
\\ 
\hline
     $\omega_{cdm}$  &
$0.139_{-0.015}^{+0.011}$
  & $0.136_{-0.014}^{+0.011}  $ &
  $0.137_{-0.014}^{+0.011}$
   \\  \hline
     $n_s$  &
$0.883_{-0.072}^{+0.076}$
  & $0.889_{-0.07}^{+0.075}$
  & $0.881_{-0.066}^{+0.07}$
   \\  
   \hline \hline 
$\Omega_m$   & $0.333_{-0.02}^{+0.019}$
& $0.329_{-0.02}^{+0.017}$
& $0.328_{-0.019}^{+0.017}$
\\ \hline
$\sigma_8$   
& $0.704_{-0.049}^{+0.044}$
& $0.711_{-0.049}^{+0.042}$ 
&  $0.699_{-0.047}^{+0.04}$\\ 
\hline
\end{tabular}
\caption{Cosmological parameter constraints from the BOSS data with the $\omega_b$ prior. $P_\ell$ denotes the $\ell=0,2,4$
moments in the range $0.01 \leq k/(\hMpc)< 0.2$, $Q_0$
is the real space power spectrum within 
$0.2 \leq k/(\hMpc)< 0.3$ (third column)
or $0.2 \leq k/(\hMpc)< 0.4$ (fourth column).
}
\label{tab:boss}
\end{center}
\end{table*}

\subsection{BOSS survey}

We apply now our method to the redshift space galaxy power spectrum
measurement of the BOSS survey~\cite{Alam:2016hwk}. 
Using the quadratic window-free estimator of Ref.~\cite{Philcox:2020vbm},
we measure the galaxy power 
spectrum multipoles of the BOSS data from four independent data chunks: 
low-z ($z=0.38$) north galactic cap (NGC), high-z ($z=0.61$) NGC, low-z south galactic cap (SGC), high-z SGC~\cite{Alam:2016hwk,Ivanov:2019pdj}. 
For each chunk we construct the likelihood as follows.
We use the full $P_0,P_2,P_4$ moments up to $\kmax=0.2~\hMpc$
and $Q_0$, estimated  with $\lmax=4$, in the ranges $0.2~\hMpc \leq k< 0.3~\hMpc$
and $0.2~\hMpc \leq k< 0.4~\hMpc$.
We do not include 
additional BAO data, as in
Refs.~\cite{Philcox:2020vvt,Chudaykin:2020ghx,Ivanov:2021zmi},
because we want to clearly 
assess the improvement from $Q_0$
w.r.t.~the usual multipoles analysis.

We fit parameters of the minimal $\Lambda$CDM model assuming 
a single massive neutrino whose mass is fixed to $0.06$ eV~\cite{Aghanim:2018eyx}, and the BBN prior on the 
baryon density $\omega_b=0.02258\pm 0.0038$.
We use the same priors on nuisance parameters as Ref.~\cite{Chudaykin:2020ghx}.
The covariance matrix for the full datavector is calculated
using the empirical estimator based on $2048$ 
Patchy mocks~\cite{Kitaura:2015uqa}. Our results 
for the joint fit of all four data chunks are displayed 
in Fig.~\ref{fig:boss} and
in Table~\ref{tab:boss},\footnote{Only the parameters that are well constrained by the data, i.e. not dominated by priors, are shown in this table.} where we show results 
from the usual
redshift-space
multipoles 
alone and
with $Q_0$,
taken at $k_{\rm max}=0.3~\hMpc$
and $0.4~\hMpc$.

In this case, the inclusion of $Q_0$ leads to somewhat marginal improvements of $\sim 10\%$, which are barely visible in
the triangle plot.
This is a result of a relatively large shot noise level of the 
BOSS galaxy sample, $\bar n^{-1}\simeq (3-5)\cdot 10^3~h^{-3}$Mpc$^{3}$. In order to illustrate this, we may analyze mocks with lower shot noise, as appropriate for the upcoming DESI survey.

\subsection{DESI-like emission line galaxy mocks}

In order to estimate the performance of our method for surveys such as Euclid and DESI, we apply it to the analysis of the mock emission line 
galaxy (ELG) catalogs from the extended Baryon Acoustic Oscillation 
Survey (eBOSS) survey~\cite{Alam:2020jvh}. These mocks simulate the clustering of the ELGs, which exhibit a weaker fingers of God signature than the BOSS LRG sample~\cite{Ivanov:2021zmi}, so the $P_\ell$ analysis is valid up to higher 
$k_{\rm max}$ in this case.
On the one hand, this factor suggests that the improvement 
from $Q_0$ may be somewhat less sizable than the improvement that we 
expect from the LRG samples. 
On the other hand, this sample has lower shot noise,
and hence the inclusion of $Q_0$ might be more beneficial here.
To understand which effect takes over, we need a quantitative 
comparison with reliable mocks, such as those recently produced using the Outer Rim simulation.

\begin{table*}[ht!]
\begin{center}
  \begin{tabular}{|c||c|c|c|} \hline
   Parameter   &  $P_{\ell}$ 
   &  $P_\ell+Q_0, (k_{\rm max}=0.3)$  
    & $P_\ell+Q_0,(k_{\rm max}=0.4)$
     \\ [0.2cm]
      \hline 
$H_0$ [km/s/Mpc]   & $71.27_{-0.43}^{+0.43}$
& $71.15_{-0.42}^{+0.41}$
& $71.09_{-0.42}^{+0.42}$\\ \hline
$\ln(10^{10}A_s)$   & $3.094_{-0.068}^{+0.066}$
& $3.111_{-0.062}^{+0.065}$
& $3.125_{-0.063}^{+0.061}$
\\ 
\hline
     $\omega_{cdm}$  &
$0.111_{-0.0048}^{+0.0043}$
  & $0.109_{-0.0045}^{+0.0039}$ 
  & $0.1079_{-0.0043}^{+0.0037}$
   \\  \hline
     $n_s$  &
$0.9896_{-0.033}^{+0.034}$
  & $0.9944_{-0.03}^{+0.032}$
  & $1.004_{-0.028}^{+0.028}$
   \\  \hline
   $b_1 $   & $1.375_{-0.034}^{+0.031}  $
& $1.364_{-0.033}^{+0.031}$ 
& $1.36_{-0.031}^{+0.03}$ 
\\ 
   \hline \hline 
$\Omega_m$   & $0.263_{-0.0081}^{+0.0071}$
& $0.2598_{-0.0076}^{+0.0064}$
& $0.2581_{-0.0069}^{+0.0064}$\\ \hline
$\sigma_8$   
& $0.8185_{-0.02}^{+0.019}$
& $0.8165_{-0.017}^{+0.017}$
&$0.8198_{-0.017}^{+0.016}$ \\ 
\hline
\end{tabular}
\caption{Constraints from the analysis of the Outer Rim mock data with the $\omega_b$ prior. 
We show only the parameters that are 
well constrained by the data. 
For $P_\ell$ the data cut is $k_{\rm max}=0.2~\hMpc$ 
in all analyses. 
For $Q_0$ we use the ranges $ 0.2~\hMpc\leq k<0.3~\hMpc$ (middle column) and $ 0.2~\hMpc\leq k<0.4~\hMpc$ (right column).
}
\label{tab:elg}
\end{center}
\end{table*}
\begin{figure}[ht!]
\centering
 \includegraphics[width=\textwidth]{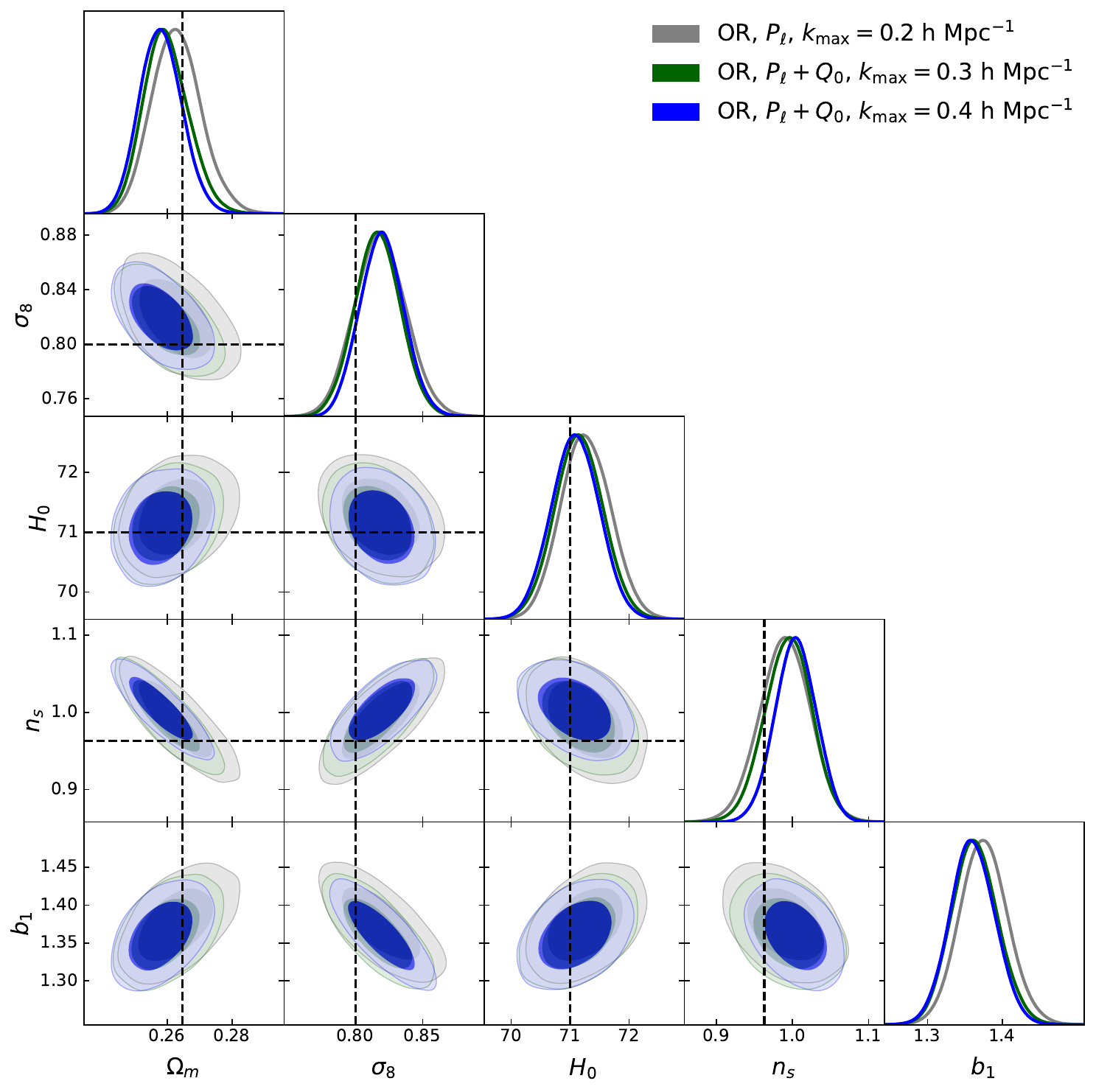}
    \caption{
Posteriors from the cosmological analysis of the 
    Outer Rim (OR) emission line galaxy mock power spectrum measurements.
    }
    \label{fig:or}
\end{figure}

These eBOSS ELG mocks are based on the Outer Rim dark matter simulation~\cite{Heitmann:2019ytn}, which were populated with ELG mock galaxies according 
to the eBOSS ELG clustering measurements~\cite{Avila:2020rmp}. 
We use the HOD-3 mock catalogs at $z=0.865$.
We combine the 27 publicly available sub-boxes into 
one large box from which we measure the mock redshift space power spectrum multipoles.\footnote{The public data on ELG mocks (based on the Outer Rim 
snapshots) is given in the form of subcatalogs 
extracted 27 nonoverlapping sub-boxes, which were cut from the original 
Outer Rim box. In the previous version of this paper, we measured the 
power spectrum from each sub-box, incorrectly assuming periodic 
boundary conditions. This has generated a bias in the $\Omega_m$ recovery.
The bias disappears when the power spectrum is measured from the cumulative catalog
produced by a proper combination of the sub-boxes, as presented above.}
The mocks have the following fiducial 
$\L$CDM
cosmology:
\be
\begin{split}
& h=0.71\,,\quad \omega_{cdm}=0.1109\,,\quad  \omega_b = 0.02258\,,\\
& n_s= 0.963\,,\quad \sigma_8 = 0.8\,, \quad M_{\rm tot}=0~\text{eV}\,.
\end{split} 
\ee
We compare three different 
analyses: fits to $\ell=0,2,4$ moments at $\kmax=0.2~\hMpc$,
fits to $\ell=0,2,4$ moments at $\kmax=0.2~\hMpc$ and $Q_0$ 
for $0.2~\hMpc\leq k<0.3~\hMpc$, and fits to 
$\ell=0,2,4$ moments at $\kmax=0.2~\hMpc$ and $Q_0$ 
for $0.2~\hMpc\leq k<0.4~\hMpc$. 
We compute the covariance in the Gaussian approximation using 
the true shot noise value $\bar n^{-1}\simeq 500$ $h^{-3}$Mpc$^{3}$
and the total volume of $V=27~h^{-3}$Gpc$^3$,
similar to the DESI ELG volume~\cite{Aghamousa:2016zmz}. 
We use the same priors on nuisance parameters as in Ref.~\cite{Ivanov:2021zmi},
but vary the spectral index $n_s$ in the fit in addition to 
$h$, $\Omega_m$ and $A_s$. 
The physical baryon density 
$\omega_b$ is fixed to the fiducial value of the simulation.

Our results are shown in Fig.~\ref{fig:or} and in Table~\ref{tab:elg}. 
First, all true cosmological parameters are 
recovered within $68\%$ confidence limits.
Second, we see that the inclusion 
of $Q_0$ shrinks the one-dimensional marginalized contours
for $\Omega_m$ and $n_s$ by $\sim 20\%$.
Third, the
posteriors do not significantly shrink 
when the data cut for $Q_0$ is increased from $0.3~\hMpc$
to $0.4~\hMpc$. 
This implies that cosmological information
in the real space power spectrum is limited
even for low shot noise samples.

All in all, we see that the improvement 
from $Q_0$
in the case of DESI-like mocks
with high number density is 
quite significant. 
Therefore, the $Q_0$ statistic can be an important 
statistic for future surveys.

\section{Conclusions}
\label{sec:disc}

In this paper, we have proposed a new statistic, dubbed $Q_0$, which acts as a proxy for the real space power spectrum, and can be used to mitigate the impact of fingers of God.
This can be easily constructed
from the conventional redshift space power spectrum Legendre multipoles.
We have shown how to perform such a reconstruction 
for an arbitrary survey and systematically include 
the information from higher-order Legendre multipoles
if they carry non-negligible signal. 
Using our approach, $Q_0$ and its covariance matrix 
can be trivially computed from theory or mock catalogs,
and included in the analysis at negligible extra cost. 
We have shown that the addition of $Q_0$ leads to notable improvements
on cosmological constraints from mock catalogs, the amplitude of which varies within $(10-100)\%$ depending 
on survey characteristics, the choice of parameters, and priors in a particular analysis.

It is useful to compare $Q_0$ to the two-dimensional redshift space power spectrum $P(k,\mu)$. 
In terms of the signal-to-noise ratio, at $k_{\rm max}=0.3~\hMpc$, the transverse moment $Q_0$ 
contains the same signal as $P(k,\mu)$ in the range $|\mu| \in [0,0.3]$. 
Thus, we expect information gains from $Q_0$ to be roughly equivalent the corresponding $\mu$-wedge.
Given that the remaining $\mu$-modes
are quite sensitive to fingers-of-God, we expect that the $|\mu|$-range $[0.3,1]$ 
contains very little, if any, viable cosmological information.

Crucially, 
$Q_0$ is more economic than $P(k,\mu)$, as it captures all relevant cosmological
information in a relatively condensed datavector. This allows us to reduce the dimensionality of the
total datavector compared to the $P(k,\mu)$ case; an effort which is of use if one wishes to avoid sampling noise 
biases if the covariance matrix is estimated from mock catalogs~\cite{Percival:2013sga}.
Alternative ways of avoiding this issue include analytic covariance matrix calculation \cite{Wadekar:2019rdu,Wadekar:2020hax} or subspace-projection techniques~\cite{Philcox:2020zyp}.

The information gain from the addition of $Q_0$ for future surveys depends on several factors. 
First and foremost, it is dependent on the strength of FoG. 
The effect is large for the BOSS-like luminous red galaxies~\cite{Alam:2016hwk,Ivanov:2019pdj}
and the bright galaxy sample to be observed by DESI~\cite{Aghamousa:2016zmz}, though less so for emission line galaxies. Hence, we expect $Q_0$
to be particularly useful for the former galaxy selections.

The second factor determining the usefulness of $Q_0$ is the particular theoretical model chosen
to fit the data, and adopted priors. 
We have found that within $\nu\Lambda$CDM the improvement from $Q_0$ increases when less restrictive priors are used and more free parameters are kept in the fit. 
Therefore, we expect even more information gain for models beyond $\Lambda$CDM, e.g. for the early dark energy scenario (see e.g.~\cite{Ivanov:2020ril} and references therein),
models with neutrino masses and additional relativistic degrees of freedom~\cite{Ivanov:2019hqk}, 
axion dark matter cosmologies~\cite{Lague:2021frh}, 
or dynamical dark energy models~\cite{DAmico:2020kxu,Chudaykin:2020ghx}.

Our work can be extended in several ways. 
First, the transverse modes measured in a realistic survey 
can be contaminated by systematics~\cite{Hand:2017irw,Ivanov:2021zmi}
so it is important to study to what extend this systematics can be 
mitigated in a realistic survey.
Second, it would be interesting to see how much $Q_0$ can improve the constraints in combination 
with other techniques, such as the bispectrum and the BAO post-reconstruction information.
This study can be performed for different tracers and within different cosmological models.
Finally, it will be interesting to work out an extension 
of our formalism to higher order statistics.
We leave these research directions for future work.

\section*{Acknowledgments}

We thank the anonymous referee for pointing out to a potential issue with the fit 
to the Outer Rim eBOSS ELG mocks,  
which motivated us 
to revisit our analysis
and to eventually fix the error that was
causing the bias in the fit.
We thank Marcel Schmittfull for his collaboration at the initial stage of this 
project.
The work of 
MI has been supported by NASA through the NASA Hubble Fellowship grant \#HST-HF2-51483.001-A awarded by the Space Telescope Science Institute, which is operated by the Association of Universities for Research in Astronomy, Incorporated, under NASA contract NAS5-26555. OP thanks the Simons Foundation for additional support. 
This work was supported in part by MEXT/JSPS KAKENHI Grant Number JP19H00677, JP20H05861 and JP21H01081.
We also acknowledge financial support from Japan Science and Technology Agency (JST) AIP Acceleration Research Grant Number JP20317829.
The simulation data analysis was performed partly on Cray XC50 at Center for Computational Astrophysics, National Astronomical Observatory of Japan.

\appendix 

\section{General relation between moments and multipoles}
\label{sec:gen}

The general relationship between power spectrum multipoles $P_{2n}$ and moments $Q_{2m}$ can be derived as follows.
The power spectrum $P(k,\mu)$ can be represented by an expansion in even Legendre polynomials ${\cal L}_{2n}$ or in even powers of $\mu$:
\begin{align}
    P(k,\mu) &= \sum_{n=0}^\infty P_{2n}(k){\cal L}_{2n}(\mu) 
    \label{eq:LegSum}
    \\
    &= \sum_{m=0}^\infty Q_{2m}(k)\mu^{2m}\;.
    \label{eq:MuPowSum}
\end{align}
They are related by 
\begin{align}
    P_{2n}(k) &= \frac{4n+1}{2}\int_{-1}^1\mathrm{d} \mu\,P(k,\mu){\cal L}_{2n}(\mu)\\
    &= \sum_{m=n}^\infty M_{nm}Q_{2m}(k),
\end{align}
where $M$ is an upper triangular matrix given by
\begin{align}
    M_{nm} = 
    \begin{cases}
    \frac{(4n+1)(2m)!}{2^{m-n}(m-n)!(2n+2m+1)!!}, \quad m\ge n,\\
    0, \quad \mbox{else}\;.
    \end{cases}
\end{align}
This follows by expressing powers of $\mu$ in terms of Legendre polynomials.
If  Eqs.~(\ref{eq:LegSum}) and (\ref{eq:MuPowSum}) can be truncated at a finite $n_\mathrm{max}=m_\mathrm{max}$, then the equations relating moments and multipoles are a finite linear system of equations, and $M$ is a $n_\mathrm{max} \times n_\mathrm{max}$ matrix.
Under that assumption the moments in terms of multipoles are 
\begin{align}
    \mathbf{Q}(k) = M^{-1} \mathbf{P}(k)
\end{align}
or explicitly
\begin{align}
    Q_{2m}(k) = \sum_{n=m}^{n_\mathrm{max}} (M^{-1})_{mn} P_{2n}(k)\;.
\end{align}
In particular, the $\mu^0$ part of $P(k,\mu)$ is a sum over all nonzero multipoles,
\begin{align}
    Q_{0}(k) = \sum_{n=0}^{n_\mathrm{max}} (M^{-1})_{0n} P_{2n}(k)\;.
\end{align}
As discussed in the main text, care must be taken when the measured power spectrum multipoles are noisy.

\bibliographystyle{JHEP}
\bibliography{short}

\end{document}